\documentclass[10pt,twocolumn,letterpaper]{article}

\usepackage[pagenumbers]{iccv}      
\usepackage{tikz}
\usepackage{braket}
\usepackage{lipsum}
\usepackage{xcolor}
\usepackage{amsmath}
\usepackage{balance}
\usepackage{environ}
\usepackage{physics}
\usepackage{colortbl}
\usepackage{enumitem}
\usepackage{amsfonts}
\usepackage{booktabs}
\usepackage{graphicx}
\usepackage{textcomp}
\usepackage{algorithm}
\usepackage{algpseudocode}
\usepackage{arydshln}
\usepackage[normalem]{ulem}

\newcommand{\sol}{\textsc{ResQ}}

\definecolor{iccvblue}{rgb}{0.21,0.49,0.74}
\usepackage[pagebackref,breaklinks,colorlinks,allcolors=iccvblue]{hyperref}

\def\paperID{5931} 
\def\confName{ICCV}
\def\confYear{2025}

\title{\sol{}:  A Novel Framework to Implement Residual Neural \\Networks on Analog Rydberg Atom Quantum Computers}

\author{Nicholas S. DiBrita\\
Rice University\\
Houston, USA\\
{\tt\small nd52@rice.edu}
\and
Jason Han\\
Rice University\\
Houston, USA\\
{\tt\small jh146@rice.edu}
\and
Tirthak Patel\\
Rice University\\
Houston, USA\\
{\tt\small tp53@rice.edu}
}

\begin{document}

\maketitle

\begin{abstract}

Research in quantum machine learning has recently proliferated due to the potential of quantum computing to accelerate machine learning. An area of machine learning that has not yet been explored is neural ordinary differential equation (neural ODE) based residual neural networks (ResNets), which aim to improve the effectiveness of neural networks using the principles of ordinary differential equations. In this work, we present our insights about why analog Rydberg atom quantum computers are especially well-suited for ResNets. We also introduce \sol{}, a novel framework to optimize the dynamics of Rydberg atom quantum computers to solve classification problems in machine learning using analog quantum neural ODEs.

\end{abstract}

\section{Introduction}
\label{sec:introduction}

The recent advancements in quantum hardware have driven the experimental demonstration of many theoretical applications in quantum computing, especially within quantum machine learning (QML)~\cite{huang2021power,biamonte2017quantum,huang2023learning,cerezo2022challenges}. This progress has spurred significant interest in implementing machine learning (ML) models like convolutional neural networks (CNNs)~\cite{rajesh2021quantum,silver2023sliq}, generative adversarial networks (GANs)~\cite{dibrita2024recon,silver2023mosaiq}, reinforcement learning~\cite{dong2008quantum,li2020quantum}, and quantum reservoir computing~\cite{mujal2021opportunities,kornjavca2024large}, largely through digital, gate-based quantum computing paradigms. Digital quantum gates allow for straightforward parameterization: certain gates depend on scalar parameters (an example being rotation gates, which are parameterized by an angle), which can be trained analogously to the weights in a neural network, enabling neural-like quantum circuits.

\begin{figure}[t]
    \centering
    \includegraphics[width=0.99\columnwidth]{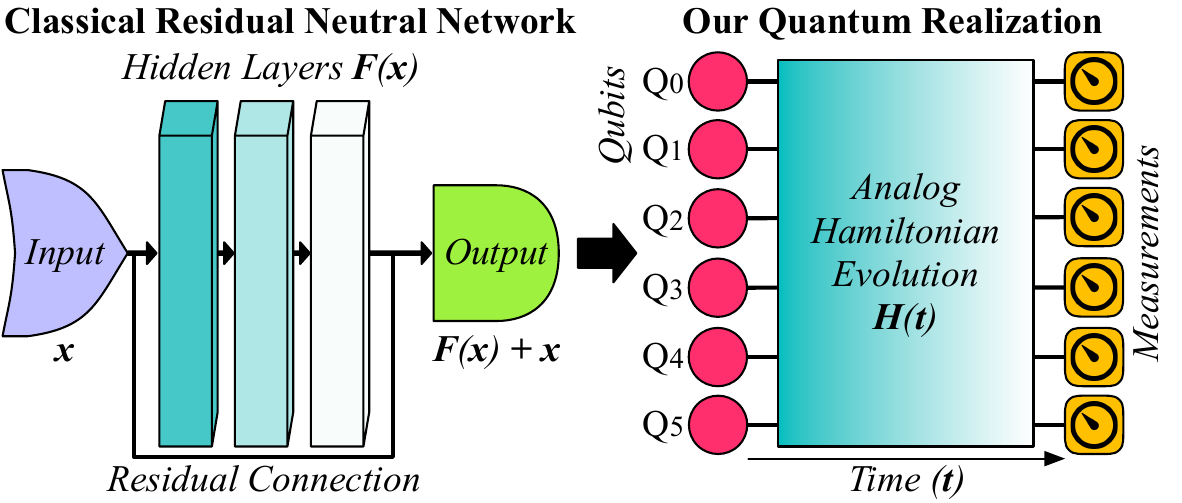}
    \vspace{1mm}
    \hrule
    \vspace{1mm}
    \caption{High-level representation of \sol{}'s analog quantum implementation of the classical residual neural network (ResNet).}
    \label{fig:residual}
\end{figure}

However, this paradigm presents limitations when extending QML to residual neural networks (ResNets). ResNets consist of layers with skip connections so that the output of a given residual block $F(x) + x$ is the sum of the input $x$ and the hidden layer output $F(x)$ (Figure ~\ref{fig:residual}). These residual blocks can help circumvent issues that arise in training via gradient-based methods~\cite{he2016deep}. Most near-term QML architectures do not have the ability to perform this skip operation, as quantum gates only allow for unitary linear transformations of the model's state.

To overcome this, we base our approach on techniques that use ordinary differential equations (ODEs) to implement residual connections~\cite{chen2018neural}. Referred to as neural ODEs, these models are parameterized to capture long-term dependencies in data. Neural ODEs involve a more complex interaction of parameters that are inherently dynamic and continuously evolving~\cite{chen2018neural}, posing challenges for the standard gate-based approach~\cite{kashif2024resqunns,killoran2019continuous}. This opens an avenue for exploring analog quantum systems, specifically Rydberg atom-based quantum computers, as a promising alternative for implementing ResNets due to their unique capacity to natively model dynamical systems (Figure ~\ref{fig:residual}).

With this insight, \textbf{in this work, we introduce \sol{}\footnote{\sol{} is published in the Proceedings of the IEEE International Conference on Computer Vision (ICCV), 2025.}, the first work to leverage Rydberg-atom quantum computers to implement ResNets for classification tasks to leverage the advantages of quantum computing.} Neural ODEs enable a constant-execution-time formulation of ResNets regardless of problem size, which aligns naturally with analog quantum computers, unlike digital quantum and classical architectures. Rydberg atom systems, in particular, are currently the only hardware that supports continuous-time Hamiltonian evolution and local/global control, making them uniquely suitable for this goal.

\sol{} parameterizes the system's Hamiltonian dynamics to encode input features and training parameters, offering a scalable and flexible platform for QML. By harnessing the globally addressed Hamiltonian in the Rydberg system, as well as the locally addressable detuning Hamiltonian, \sol{} enables precise control over atom states and interactions, allowing us to fine-tune the system to encode complex data inputs and train ResNet models effectively. This approach enhances the strengths of analog Rydberg systems to enable a differential equation-based residual structure.

\vspace{2mm}

\begin{figure}[t]
    \centering
    \includegraphics[width=0.99\columnwidth]{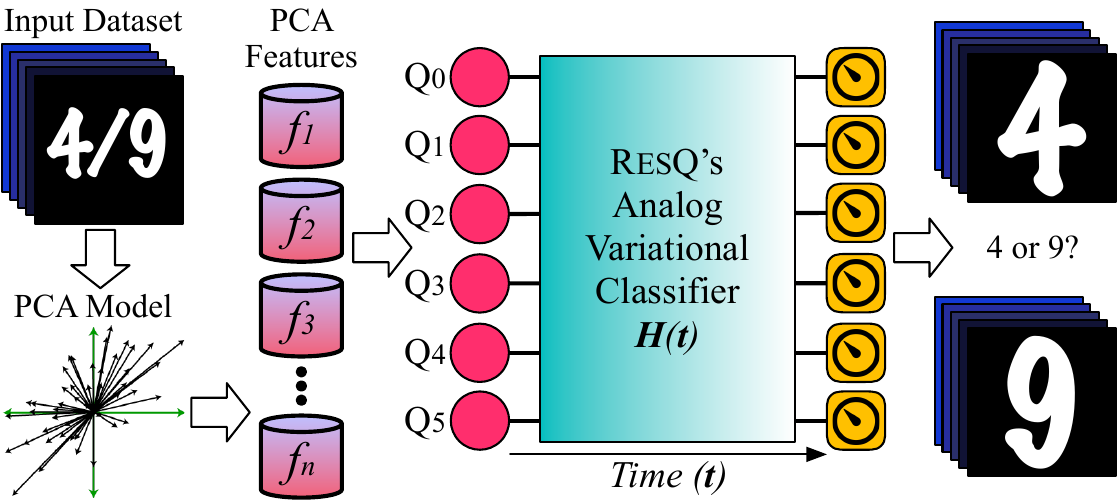}
    \vspace{1mm}
    \hrule
    \vspace{1mm}
    \caption{Representation of \sol{}'s classification workflow for the example of classifying 4 vs. 9 in the MNIST dataset~\cite{mnist}. Note that \sol{} uses the measurement of all the qubits.}
    \label{fig:resq}
\end{figure}

\noindent \textbf{Our contributions are summarized as follows:}

\begin{itemize}[leftmargin=*]

    \item We derive and demonstrate the natural alignment between residual networks and the analog dynamics of Rydberg atom computers, underscoring their mutual suitability.

    \vspace{2mm}

    \item Through the design of \sol{}, we establish an open-source methodology to realize ResNets; using piece-wise parameterization, \sol{} breaks down input data and model parameters into controllable intervals across the Hamiltonian dynamics.

    \vspace{2mm}

    \item \sol{} also facilitates the linear scaling of features with qubit count using local detuning coupling strength for parameterization, a significant advancement in ensuring that ResNets can handle larger datasets without increasing the circuit depth (length of the critical path). 

    \vspace{2mm}

    \item Using \sol{}, we implement ResNet-based classification on the MNIST and FashionMNIST datasets and healthcare datasets, conducting evaluation across ideal and noisy simulations based on QuEra's Aquila 256-qubit quantum computer. We find that \sol{} can outperform similarly sized classical models, accomplishing a 56\% and 57\% improvement over simple feedforward networks and simple ResNets, respectively. It also achieves a 36\% improvement over classical neural ODE classifiers.

    \vspace{2mm}

    \item We explore various atom configurations, including chain, ring, square, and triangular lattices, and empirically demonstrate their scaling and impact on atomic interaction strengths and classification accuracy.

    \vspace{2mm}
    
    \item Lastly, we perform a real-computer evaluation on the 256-qubit QuEra Aquila analog Rydberg-atom quantum computer to show the minimal impact of hardware noise due to \sol{}'s noise-resilient design.

    \vspace{2mm}

    \item The experimental code and dataset related to this work are open-sourced at: {\small \url{https://github.com/positivetechnologylab/ResQ}}.
    
\end{itemize}
\section{Brief and Relevant Background}
\label{sec:background}

\begin{figure*}[t]
    \centering
    \subfloat[\sol{}'s Atom Grid Configurations]{\includegraphics[width=0.79\textwidth]{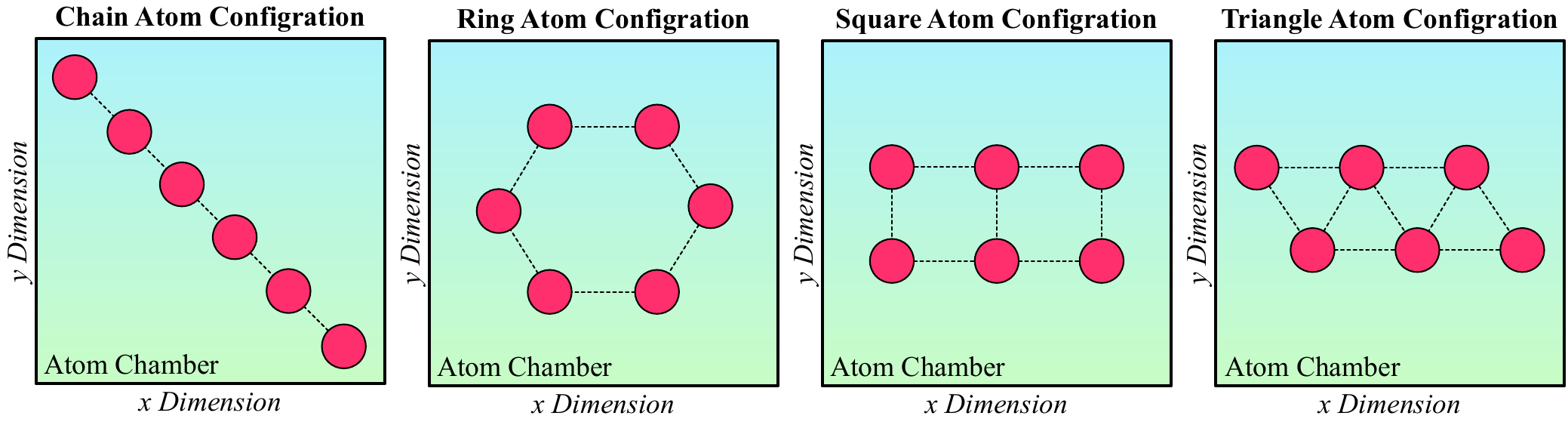}}
    \hfill
    \subfloat[\sol{}'s Scaling of Grid]{\includegraphics[width=0.195\textwidth]{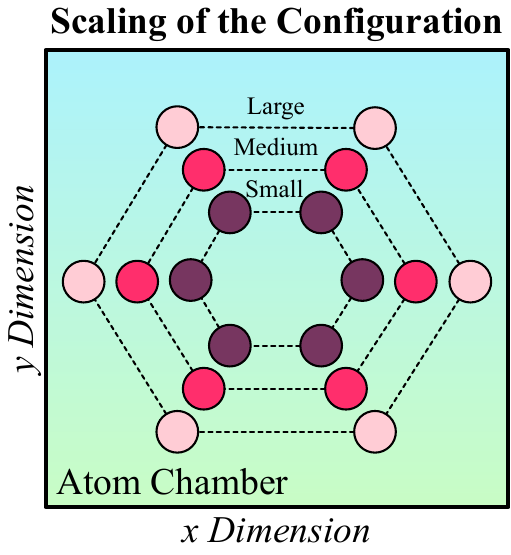}}
    \vspace{1mm}
    \hrule
    \vspace{-2.5mm}
    \caption{(a) \sol{} explores different atom configuration grids: chain, ring, square, and triangle. (b) \sol{} explores different grid scales.}
    \label{fig:layouts}
\end{figure*}

\subsection{Quantum Computing and Quantum Machine Learning Fundamentals}

Quantum computing leverages the principles of quantum mechanics to enable new forms of computation. Unlike classical bits that take binary values (0 or 1), quantum bits, or \textbf{qubits}, can exist in a \textbf{superposition} of both states. A qubit's state is written as $|\psi\rangle = \alpha |0\rangle + \beta |1\rangle$, where $\alpha$ and $\beta$ are complex amplitudes such that $|\alpha|^2 + |\beta|^2 = 1$. This can be viewed as a unit vector in a two-dimensional complex vector space, analogous to how feature vectors represent images or embeddings in computer vision. Upon measurement, the qubit collapses to one of the basis states: $|0\rangle$ with probability $|\alpha|^2$ or $|1\rangle$ with probability $|\beta|^2$.

In addition to superposition, qubits can exhibit \textbf{entanglement}, a quantum phenomenon where the state of one qubit becomes dependent on the state of another, even across long distances. This introduces complex correlations between qubits that cannot be described by classical joint distributions. When multiple qubits are measured after a quantum computation, the result is a \textbf{probability distribution} over all possible bitstrings, akin to a probabilistic output layer in a neural network~\cite{patel2020disq,ash2019qure,bhattacharjee2019muqut}.

These properties allow quantum computers to represent and explore vast solution spaces in parallel, offering potential speedups for tasks involving high-dimensional search, optimization, and sampling. QML sits at the intersection of quantum computing and machine learning, with the goal of accelerating or enhancing ML models through quantum resources. Most existing QML research builds on digital, gate-based quantum systems, where qubits are transformed by applying discrete gates, such as the rotation gate $R_x(\theta) = \exp(-i \theta X / 2)$, with $X$ being the Pauli-X matrix~\cite{patel2022optic,wang2022quantumnat,ranjan2024proximl}. This is loosely analogous to applying a sequence of nonlinear functions in a deep network.

However, digital quantum circuits struggle to express models with inherently continuous dynamics, such as residual networks (ResNets), neural ODEs, or diffusion models, without decomposing them into many small discrete steps. In contrast, analog quantum systems (such as Rydberg atom arrays) offer native support for continuous time evolution, making them especially well-suited for implementing such models directly in hardware.

\subsection{Residual Neural Networks}

ResNets are deep learning models that address challenges in training deep networks, such as vanishing gradients, by incorporating \textbf{residual connections} that allow information to bypass layers~\cite{fang2021deep,he2016deep}. ResNets can be expressed in terms of ODEs to capture continuous transformations within the data~\cite{chen2018neural}. Moving to the continuous-time limit, a ResNet can be described as an ODE system $\frac{dx}{dt} = F(x, \theta)$, where $x$ is the input state and $\theta$ is the model parameter vector. 

Referred to as a neural ODE, this continuous, evolving structure makes ResNets highly effective for capturing complex, long-term dependencies in data but poses challenges for digital quantum systems, which rely on discrete gate operations~\cite{liang2021hybrid,wen2024enhancing}. Analog quantum systems, which naturally undergo continuous temporal evolution, are a more promising approach for implementing ResNets by way of neural~ODEs.

\subsection{Rydberg Atom Quantum Computers and Analog Quantum Computers}

Rydberg atom quantum computers use neutral atoms as qubits, taking advantage of atoms in high-energy \textbf{Rydberg states}. In these states, atoms exhibit expanded electron orbits and elevated energy levels, which result in strong, distance-dependent interactions with nearby atoms. The atoms are spatially arranged in regular geometric configurations, such as one-dimensional chains or two-dimensional grids, within optical traps or lattices formed by laser interference patterns. External laser pulses manipulate each atom to act as a qubit (we use the terms ``atoms'' and ``qubits'' interchangeably in this work), with their interactions controlled by tuning excitation energy, spatial positioning, and inter-atom distances.

Analog Rydberg systems can be loosely compared to continuous-time dynamical systems like neural ODEs or residual networks (ResNets). Unlike gate-based quantum systems that apply a sequence of discrete operations, analog quantum computers evolve continuously under the influence of an external control field. This makes them especially suitable for representing smooth transformations in time, analogous to how a ResNet layer adds a learned residual function to its input (Fig.~\ref{fig:residual}). By encoding both data and model parameters directly into the evolution of the quantum system, we realize a form of continuous information processing that can be viewed as the quantum counterpart to time-evolving deep learning architectures~\cite{kornjavca2024large,wilson2024non,lu2401digital}.

In analog quantum computation, the \textbf{Hamiltonian} $H(t)$ plays the role of the system's energy function, dictating its time-dependent evolution through Schrödinger’s equation: $i\hbar \frac{d}{dt} |\Psi(t)\rangle = H(t) |\Psi(t)\rangle$, where $|\Psi(t)\rangle$ is the quantum state at time $t$ and $\hbar$ is the reduced Planck constant~\cite{dibrita2024recon}. This is conceptually similar to how the gradient in a neural network governs the update of parameters, except here, the Hamiltonian shapes the trajectory of the quantum state in a high-dimensional complex vector space. For an $N$-qubit system, $|\Psi(t)\rangle$ is a complex vector of dimension $2^N$, so the state space grows exponentially with the number of qubits, providing exponentially expressive capacity, albeit being difficult to simulate classically.

In this work, we use the 256-qubit QuEra Aquila Rydberg atom-based quantum computer (the only publicly available Rydberg atom computer)~\cite{wurtz2023aquila} to implement ResNets for classification tasks in QML. \textit{By encoding inputs into Aquila’s global and local control settings, \sol{} enables a scalable framework that aligns with ResNets’ continuous dynamics, providing a flexible and powerful platform for QML applications to run on quantum computers.} 

\section{Related and Prior Work}
\label{sec:related_work}

Recent work in QML has extended classical neural network architectures to quantum systems, focusing on residual connections to address limitations in quantum neural networks (QNNs). Wen et al.~\cite{wen2024enhancing} introduced residual connections to QNNs on gate-based systems, enhancing expressivity and helping the networks capture more complex functions. Kashif and Al-Kuwari~\cite{kashif2024resqnets} developed ResQNets to mitigate barren plateaus, preserving gradient flow in deep QNNs, but also limited to gate-based quantum systems. Nakajima~\cite{Nakajima2021} investigated Schrödinger-based neural ODEs and the use of nonlinear optics as activation functions. Banchi~\cite{banchi2021measuring} and Leng~\cite{leng2022differentiable} also explore the connection between neural ODEs and Schrödinger dynamics and develop a methodology for computing approximate gradients of these techniques using efficient methods.

Parallel to these works, neural ODEs have also been used for physics-informed machine learning to model quantum systems and their dynamics~\cite{Choi2022}. Hybrid quantum-classical models have incorporated residual structures to improve QML performance. Liang et al.~\cite{liang2021hybrid} presented a hybrid model with residual learning to boost convergence, combining classical and quantum layers for flexibility. Similarly, Abd El-Aziz et al.~\cite{abd2022quantum} used a residual approach within a quantum optimization technique for IoT, but their model also relies on classical computation. While effective, these models are not fully quantum-native.

Other approaches, such as continuous-variable QNNs introduced by Killoran et al.~\cite{killoran2019continuous}, leverage continuous-variable basis to implement neural networks, including ResNets. This work opened new avenues for continuous-variable computation but did not leverage the qubit-based Rydberg atom analog computing capabilities. These capabilities are a central part of \sol{}.

On the other hand, Rydberg atom computers have been used by~\cite{Lu_2025} for binary MNIST classification. This work uses a hybrid approach that combines digital gates with analog Hamiltonian evolution. While this leads to good performance across the different classes in ideal simulation scenarios, current commercially available Rydberg atom computers are only capable of analog functionality, meaning this technique is not applicable to available devices.

\textit{\sol{} is the first framework to implement ResNets in a fully quantum-native analog system to leverage quantum properties.} By aligning the time-evolving dynamics of Rydberg systems with the ODE-based structure of residual networks, \sol{} achieves efficient feature scaling and model depth, overcoming the limitations of gate-based QML and hybrid models. This positions \sol{} as a novel approach, advancing analog systems for complex tasks.
\section{\sol{}'s Design}
\label{sec:design}

The overview of \sol{}'s design for classification tasks is provided in Figure \ref{fig:resq}. First, datasets are preprocessed with PCA for dimensionality reduction. PCA features are rescaled so they obey physical constraints and are used as inputs to the analog Rydberg computation. The inputs and parameters are encoded in the Hamiltonian terms. The Hamiltonian is evolved in time, and at the end of the evolution, the qubits are measured, and the probability of measuring $\ket{1}$ (averaged across all qubits) gives the prediction. The Hamiltonian parameters are trained through an adaptive gradient-based technique based on cross-entropy loss.

We now describe the design in detail.

\subsection{Analog Rydberg Atom Dynamics}

\begin{figure}[t]
    \centering
    \includegraphics[width=0.99\columnwidth]{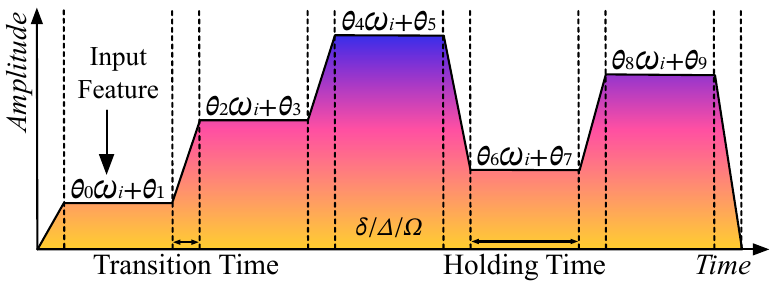}
    \vspace{1mm}
    \hrule
    \vspace{1mm}
    \caption{\sol{} divides the pulses into multiple intervals (five in this example), with two parameters per interval, including the scaling for the input feature ($\omega_i$) and the offset.}
    \label{fig:amplitudes}
\end{figure}

The globally addressed Rydberg Hamiltonian that describes the dynamics of the Rydberg atom quantum computer is:
\begin{equation}
\begin{split}
H(t) = \frac{\Omega(t)}{2} \sum_i \left( e^{i\phi(t)} |g\rangle_i \langle r|_i + e^{-i\phi(t)} |r\rangle_i \langle g|_i \right) \\
- \Delta(t) \sum_i \hat{n}_i + \sum_{i < j} \frac{C_6}{|\vec{p}_i - \vec{p}_j|^6} \hat{n}_i \hat{n}_j    
\end{split}
\end{equation}

In the above, $t$ represents time. The function $\Omega(t)$ denotes the \textit{globally addressed} Rabi frequency, which is the amplitude of the drive field applied uniformly to all qubits at time $t$. The function $\Delta(t)$ represents the detuning at time $t$, defined as the difference between the drive frequency and the natural transition frequency of the qubits. The function $\phi(t)$ denotes the relative phase of the drive field at time $t$, which determines the phase offset of the applied control signal. These three components form the global pulse.

Global addressing refers to the fact that these Hamiltonian parameters are the same across all atoms.  $|g\rangle_i$ and  $|r\rangle_i$ are the ground and excited Rydberg states of atom $i$ respectively~(~$\bra{r} = \ket{r}^\dagger$, where $\dagger$ represents the Hermitian conjugate). For all calculations, we use the convention $\ket{g} = \ket{0}, \ket{r} = \ket{1}$. $\hat{n}_i$ counts the number of Rydberg excitations for atom $i$ and is equivalent to $|r\rangle_i \langle r|_i$. $\frac{C_6}{|\vec{p}_i - \vec{p}_j|^6}$ represents the Rydberg interaction potential, with $\vec{p}_i$  being the location of the atom $i$ in the grid and $C_6 = 862690 \cross 2\pi$ MHz $\mu m^6$ being a constant.

QuEra Aquila also has locally addressed detuning, which adds the following local term to the above Hamiltonian:
\begin{equation}
    H_{local}(t) = -\delta(t)\sum_i  h_i \hat{n}_i
\end{equation}

Here, $\delta(t)$ is the local detuning at time $t$, which is the same for all atoms, and $h_i \in \left[0, 1\right]$ is a site-dependent coupling to the detuning. This site-dependent parameter allows for finer control over individual atoms than the global terms above. \sol{} leverages these time-dependent parameterizable pulses for its implementation for classification tasks, as described later in this section.

\subsection{Atom Grid Initialization}

\sol{} uses the quantum dynamics of the Rydberg atom computer to perform residual network computations. These computations are parameterized in terms of the parameters in the computer's Hamiltonian. \textit{\sol{} first sets the spatial positions of the atoms based on one of the four basic configurations from Figure \ref{fig:layouts}(a).} These configurations were selected for their ability to represent different interaction dynamics of different datasets. For instance, if a dataset has two-local feature interactions among nearest neighbors, it would benefit from the chain or ring configuration.

Each of these configurations also depends on a scaling parameter, as depicted in Figure \ref{fig:layouts}(b), which determines the size of the grid spacing and, thus, the interaction and entangling strength of the atoms -- this strength scales inversely with the distance between the atoms. In principle, the exact choice of grid configuration and spacing is application-specific; however, we generally find that moderate spacings with weak interactions perform best in practice (Section ~\ref{sec:evaluation}).

\subsection{Laser Dynamics}

Once the atoms are spatially configured, they are ready for the Hamiltonian evolution. Each laser parameter $\Omega, \Delta, \text{ and } \delta$ is specified as a time series. \sol{} uses a parameterization that consists of multiple piecewise-linear pulse intervals, shown in Figure \ref{fig:amplitudes}. \textit{A key innovation of \sol{} is to parameterize the strength of each pulse (pulse $i$) at the ``holding time'' as $\theta_j \omega_i + \theta_{j+1}$.} Here,  $\theta_j \text{ and } \theta_{j+1}$ are parameters that are learned during training, while $\omega_i$ is a single input to the model. This enables each pulse interval to be impacted by a learned scaling of the input feature and an additional offset, allowing the input feature to have a notable influence over the entire pulse duration. Note that these input features could take any form: direct image pixel features or PCA features. In our case, we embed them as PCA features to enable scaling down of the feature space.

The rest of the pulse is then formed by linear interpolation between these holding times. The number of intervals $M$ is limited by the allowed runtime of the computer and the duration of each interval. We use a hold time of $0.15\mu s$ and a transition time of $0.05\mu s$, which, for the maximum runtime of $4.0\mu s$, allows for a maximum $M=19$. This specific timing was chosen as an interval of 0.2$\mu s$ corresponds to a $\pi$-pulse at the maximum allowed Rabi frequency of Aquila, 15.8 $\frac{rad}{\mu s}$. \sol{} uses this parameterization for $\Omega, \Delta, \text{ and } \delta$, giving 3 inputs to the model and $6M$ trainable parameters. This enables us to use the pulse dynamics as tuning knobs to optimize the parameter values during training. We set $\phi=0$ for all $t$, as it was found not to meaningfully impact the results, and this simplifies the gradient calculations.

\begin{figure}[t]
    \centering
    \includegraphics[width=0.99\columnwidth]{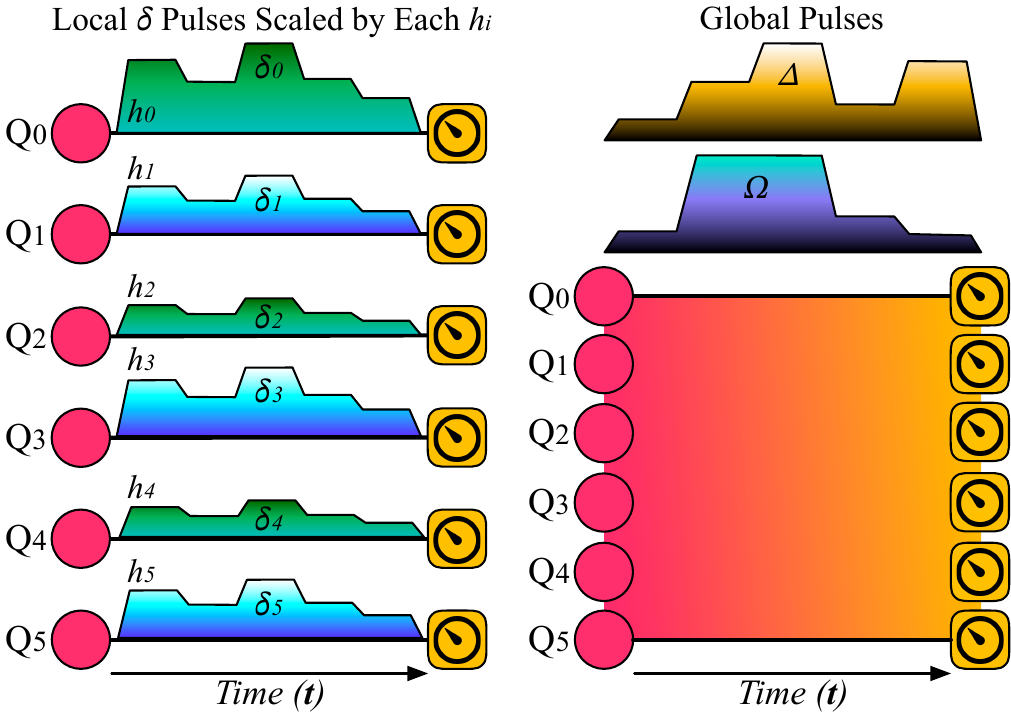}
    \vspace{1mm}
    \hrule
    \vspace{1mm}
    \caption{Pluses used by \sol{}. All local detuning pulses ($\delta_i$) have the same pulse shape, which is individually scaled for each qubit -- per the $h_i$ value. Even qubit $h_i$'s encode the input features, and odd qubit $ h_i$'s encode the parameters.}
    \label{fig:pulses}
\end{figure}

In addition to the atom positions and pulse shapes, we also need to set the site-dependent coupling $h_i$ for the local detuning. \sol{} uses half of these couplings as input and the other half as trainable parameters in an alternating fashion, as shown in Figure \ref{fig:pulses}. This allows us to use local addressing to embed certain data features into the computation while still maintaining some local control that is tunable. For $N$ qubits, this adds an additional $\frac{N}{2}$ inputs to the model and $\frac{N}{2}$ trainable parameters. \textit{In total, the design of \sol{} enables $3 + \frac{N}{2}$ input features and $6M + \frac{N}{2}$ residual network parameters, which allows it to directly increase the size of the parameter space or feature space by increasing the number of atoms $N$. The number of trainable parameters can be further increased by increasing the number of pulse intervals $M$.}

All parameters are initialized to be 1.0, as this prevents any one term in the Hamiltonian from dominating the evolution at the start of training.

\subsection{Dimension Reduction and Scaling}

Although the allowed number of input features is quite large for the 256-qubit Aquila (as the number of inputs scales linearly with the number of qubits in our design), it is still desirable to reduce the size of the input data, especially when running on today's noisy quantum computers, where noise is proportional to the number of qubits. It is also desirable to reduce the input size (i.e., the number of qubits) to allow for the simulation of quantum dynamics classically.

\sol{} uses Principal Component Analysis (PCA) for this purpose, as it is relatively quick to fit datasets and can effectively account for most of the variance using only a small subset of the principal components. The resulting PCA features are then MinMax scaled so that all of the Hamiltonian parameters are within the allowed ranges dictated by the physical hardware constraints. For the pulse inputs $\omega_i$'s of $\Omega  \text{ and }  \Delta$, \sol{} sets a fixed range of $[\frac{\pi}{2}, 2\pi]$, while $\delta$ is the same range but negative due to additional hardware restrictions. \textit{The key insight here is that having a non-zero lower bound forces some dynamics to be present in the computation and prevents the system from always remaining in the initial $\ket{00...0}$ state during training.}

The coupling inputs $h_i$ are simply scaled to be within the hardware allowed $[0, 1]$. Since PCA features are ordered by the proportion of variance they explain, it is important that we assign the high-variance features to inputs that will most impact the computation. \textit{We design \sol{} to assign the first PCA features to the pulse inputs of the globally addressed terms ($\Omega \text{ and } \Delta$) and assign less important features to the local terms, as this was found to give the best performance.}

\subsection{Measurement}

After the Hamiltonian evolution is completed, a measurement is performed on all qubits. Depending on the output state of the computer, each qubit will return either 0 or 1 based on the coefficients of the state vector, as mentioned in Section ~\ref{sec:background}. The computation is repeated a number of times so that we can estimate the probability of measuring each qubit to be in the $\ket{1}$ state. We then average these probabilities and use this final result as the soft label output from our model. Rounding is applied afterward to extract hard labels -- the probability of measuring $\ket{1}$ state of less than 0.5 is assigned to one class, and more than 0.5 is assigned to the other. \textit{We select this method over the conventional QML method of measuring only one qubit as it allows for the various input features to have some influence on the final state}, even when atoms in different parts of the computer only weakly interact during the computation.

\subsection{Loss Function and Gradient Calculations}

We train the residual network with respect to binary cross-entropy loss. To calculate the gradients, we adapt the stochastic pulse gradient method~\cite{banchi2021measuring}~\cite{leng2022differentiable}, which provides an unbiased stochastic estimate for the gradient of analog quantum programs. A full write-up of the procedure can be found in Algorithm 1 in~\cite{leng2022differentiable}. To compute the gradient, we repeatedly perform the same analog routine as the reference computation, with additional qubit rotations randomly inserted at different time points during the evolution. These repeated executions are then used to estimate an integral, the solution of which gives the gradient to that analog program. This estimate can be improved by increasing the number of sampled points. However, this increases the number of quantum hardware executions or classical simulations required for each gradient update. \textit{\sol{} uses 20 samples per gradient evaluation, as this empirically provides a balance between accuracy and simulation time.}

\vspace{2mm}

\noindent With that, we conclude the design description of \sol{}. Next, we describe its evaluation methodology.
\section{Experimental Methodology}
\label{sec:methodology}

\begin{figure}[t]
    \centering
    
    \includegraphics[width=0.99\linewidth]{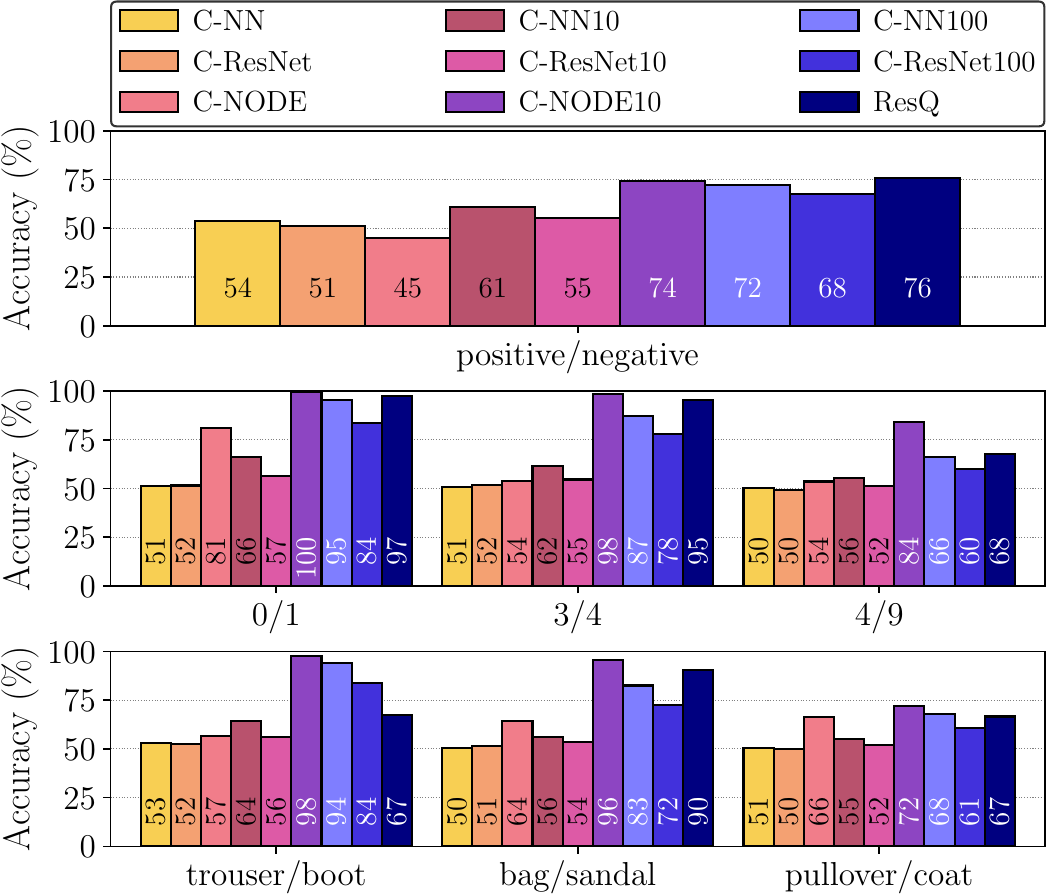}
    \vspace{1mm}
    \hrule
    \vspace{1mm}
    \caption{Accuracy scores for the Pima Indian Diabetes dataset (top), 3 different MNIST classification tasks (middle), and 3 different FashionMNIST classification tasks (bottom). \sol{} outperforms various classical neural network techniques with similar parameter counts when trained for the same number of iterations. Additionally, it typically performs better than or comparable to models with 10 or 100 times as many parameters. }
    \label{fig:accuracycomparision}
\end{figure}

\noindent\textbf{Software and Simulation Setup.} All training experiments are done using classical simulation. We use $N = 4$ atoms for all parts of the evaluation, as this gives a reasonable trade-off between performance and simulation time. All implementation and analysis code is written in the Julia language. Classical simulation of the analog quantum computation is performed using Bloqade.jl. Classical neural networks are implemented using Flux.jl, while classical neural ODEs are implemented in DiffEqFlux.jl. Optimization of various models is performed with the Adam optimizer, which is implemented in Optim.jl. Gradient calculations are performed using the stochastic pulse gradient method, which we implemented in Julia using operations from Bloqade.jl and Yao.jl. All simulation-based experiments are run on our local cluster, with nodes comprised of the AMD EPYC 7702P 64-Core processor. We spawn virtual machines consisting of 8 cores and 32 GB of memory, running Ubuntu 20.04.6 LTS.


\vspace{2mm}

 \noindent\textbf{Quantum Hardware.} Real hardware inference runs are completed on QuEra's Aquila device, an analog Rydberg atom computer containing 256 rubidium atoms as qubits. We submit quantum tasks to Aquila through Amazon Braket~\cite{aws}. Currently, Aquila allows users to customize the 2D spatial placement of the atoms, as well as specify the global laser parameters $\Omega(t)$, $\Delta(t)$, and $\phi(t)$~\cite{wurtz2023aquila}. Local detuning is also available; however, support for this is currently only experimental. We make use of this experimental feature, as well as the global laser parameters and the spatial configuration. As with the simulation, all hardware runs use $N = 4$ atoms, with parameters and spacing determined from previous training done using simulation. Each inference run is submitted as a separate quantum task, which is repeated 1000 times in order to get good measurement statistics of the resulting qubit probability distributions.

\vspace{2mm}

\noindent\textbf{Comparative Classical Techniques.} We compare \sol{} against classical neural architectures of different sizes, including feed-forward networks (denoted as C-NN), residual networks (denoted as C-ResNet), and a neural ODE model (denoted as C-NODE). We test each of these models with varying parameter counts (approximately $\times$1$, \times$10, or $\times$100 the number of parameters \sol{} has). The smallest C-NN and ResNet models consist of one hidden layer with ReLU activation and softmax applied to the output. The two larger-sized C-NN and ResNet models have 4 hidden layers and, again, a softmax output. In ResNet, the residual connection is formed around all hidden layers. The neural ODE has its derivative function expressed as the output of a feed-forward network, which itself has a single hidden layer with ReLU activation. The output of the numerical ODE solution is then fed into one final linear layer, as discussed in~\cite{dupont2019augmented}.

For every task, we train all models using gradient-based methods. We use the same training data as \sol{}, in order to make a fair comparison to the training done for \sol{}. No minibatching is used, and the same number of iterations (75) is used for all quantum and classical techniques. Note: we could not compare to any competitive quantum methods as \sol{} is the first work in this domain.

\vspace{2mm}

\noindent\textbf{Benchmarks and Datasets.} We evaluate \sol's performance on several binary classification tasks, using data from MNIST~\cite{mnist}, FashionMNIST~\cite{xiao2017fashion}, and the Pima Indians Diabetes Database~\cite{Pima} (PID). For each data set, we use 5 PCA features (3 for the pulses and 2 for the local coupling strengths). For MNIST tasks, pairwise classification is performed on 0 and 1 (easy), 3 and 4 (medium), and 4 and 9 (hard) to show a wide range of \sol{}'s performance. This classification difficulty has been determined empirically by prior work~\cite{ranjan2024proximl}. Similarly, for FashionMNIST, classification is performed on trousers and boots (easy), bags and sandals (medium), and pullovers and coats (hard). In the diabetes dataset, 
we compare the negative diagnoses against the positive diagnoses for binary classification. For MNIST and FashionMNIST, we randomly sampled 1000 points for the training data to reduce the number of classical simulations required during training to a manageable size. We use the same samples for training comparative classical techniques. For each classification task, we computed accuracy and the F1 score based on a separate test set.

\section{Evaluation and Analysis}
\label{sec:evaluation}

\subsection{\sol{}'s Flagship Results and Performance} 


\begin{figure}[t]
    \centering
    \includegraphics[width=0.99\linewidth]{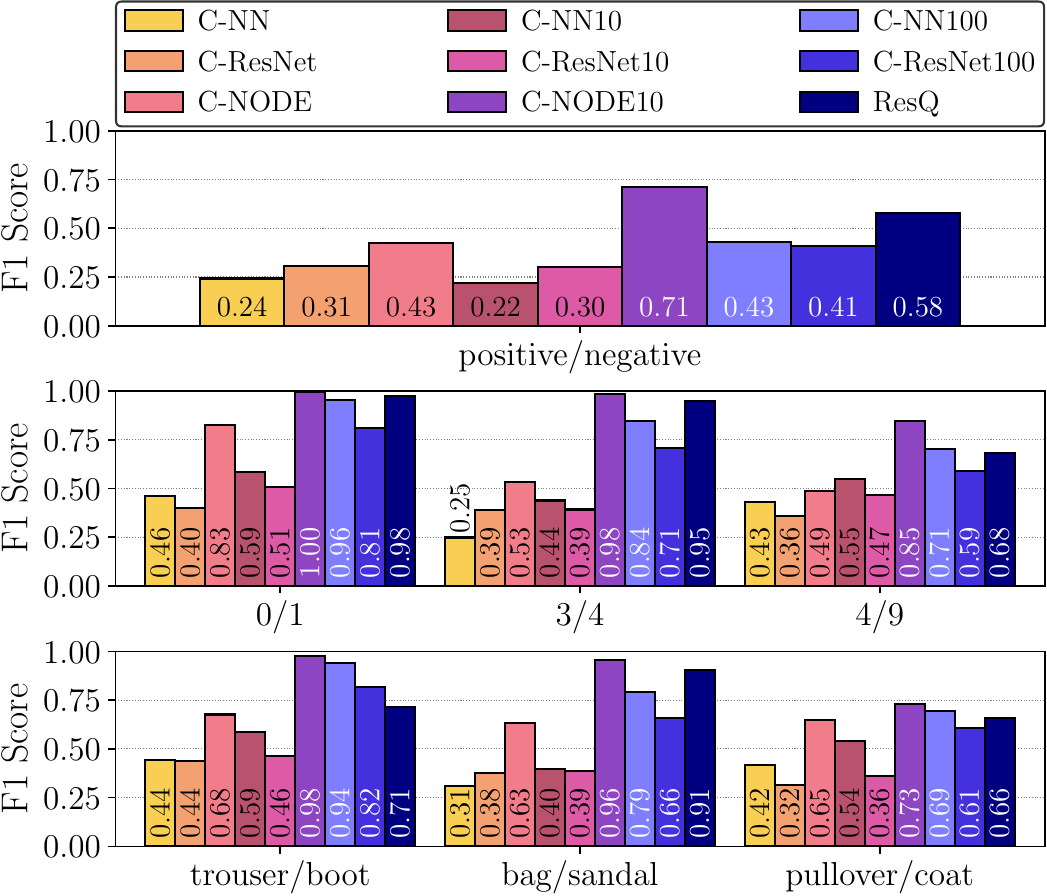}
    \vspace{1mm}
    \hrule
    \vspace{1mm}
    \caption{F1 scores for MNIST classification tasks (top) and FashionMNIST tasks (bottom). Similar to the accuracy results, \sol{} performs at a level comparable to larger models ($10\times{}$ and $100\times{}$) and outperforms models with similar parameter counts. We omit the PID database scores for brevity, but they show a similar trend, with \sol{} only being outperformed by the larger models.}
    \label{fig:f1comparision}
\end{figure}

In Figure~\ref{fig:accuracycomparision}, we compare \sol{}'s testing accuracy with a classical feedforward network, a residual network, and a classical neural ODE. We compare against models that have the same number of parameters as \sol{}, as well as models that have 10 or 100 times as many parameters.

\sol{} manages to outperform all other comparative techniques on the PID classification task accuracy, regardless of parameter count. \sol{} outperforms the 3 classical models that have a similar parameter count, showing a 52\% improvement on average. It also performs better than or comparably to models 10 and 100 times larger in parameter count. We see the same behavior in the MNIST classification tasks, with an even larger 61\% improvement over the similarly-sized classical models.

In the FashionMNIST tasks, we see a smaller but still substantial 37\% improvement over the various small classical models. In the trouser/boot classification, the larger classical models outperform \sol{} as they have a much larger parameter count. This is because \sol{} optimizes for parameter efficiency in the available analog regimes. However, in the other two tasks, \sol{} still remains competitive with its few parameters. 

Across all of the tests performed, \sol{} gives a 50\% average improvement over similarly-sized classical techniques. This includes a 56\% improvement over C-NN, a 57\% improvement over C-ResNet, and a 36\% improvement over C-NODE. It does this while maintaining good precision and recall, as shown in the F1 scores (Figure~\ref{fig:f1comparision}). For the F1 scores, \sol{} maintains its advantage over similarly sized models and can again rival larger classical models in most of the classification tasks, indicating it is capable of providing robust predictions. 

\begin{figure}[t]
    \centering
    \includegraphics[width=0.99\linewidth]{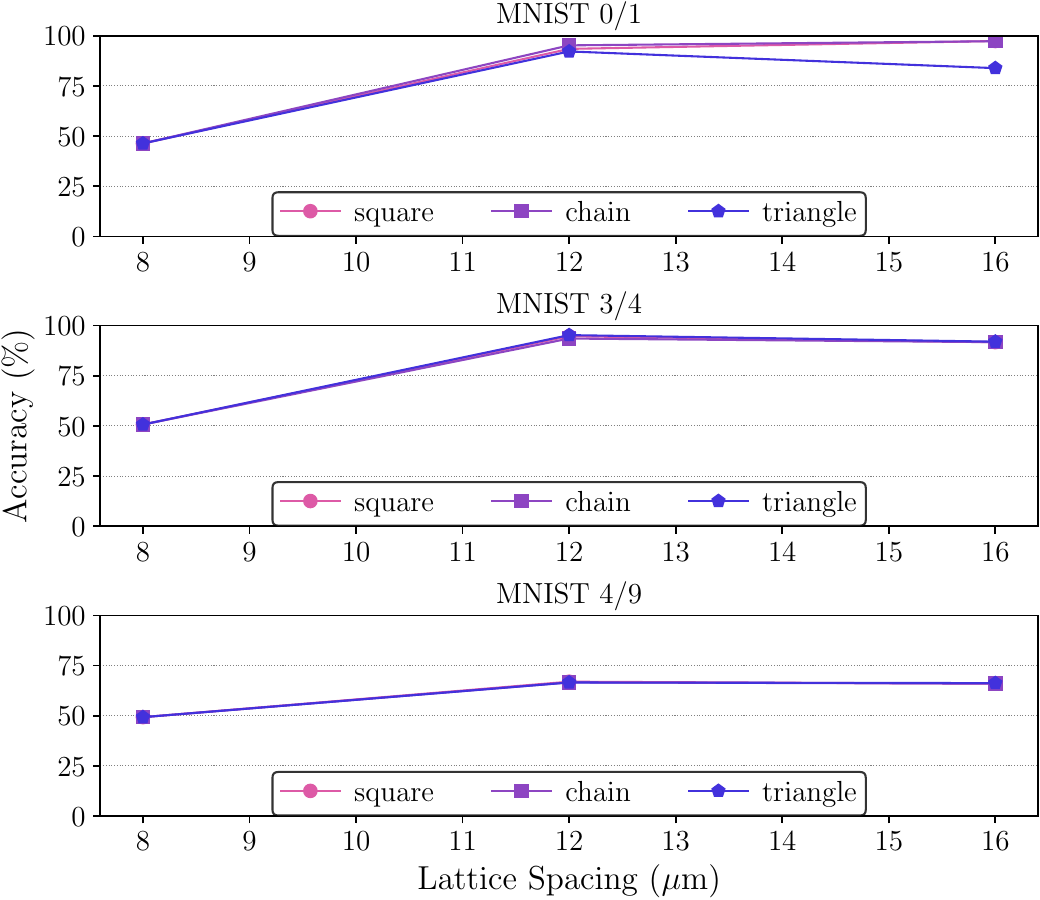}
    \vspace{1mm}
    \hrule
    \vspace{1mm}
    \caption{Performance of \sol{} for different atom configurations and spacings as tested on different pairs of MNIST classes. }
    \label{fig:spacing_acc}
\end{figure}

\subsection{Impact of Atom Configuration on \sol{}} 

To explore the effect of lattice configuration, we trained \sol{} on the benchmark datasets with three different lattice types and three different spacings. Note that for N=4, the ring lattice is identical to the square lattice and is therefore not shown for brevity. An example of this is shown for MNIST in Figure~\ref{fig:spacing_acc}. Here, we see that the different lattice types can all offer similar performance on this data set, with a slight preference given to square and chain. \sol{} also prefers a moderate amount of spacing in the lattice, with performance plateauing at spacings beyond 12$\mu m$. We found this behavior consistent across our dataset benchmarks, which indicates that weak entangling interactions are desirable for most tasks, but strongly interacting systems may not be due to the difficulty of training them. We observed similar trends for other datasets and classes.

\begin{figure}
    \centering
    \includegraphics[width=0.99\linewidth]{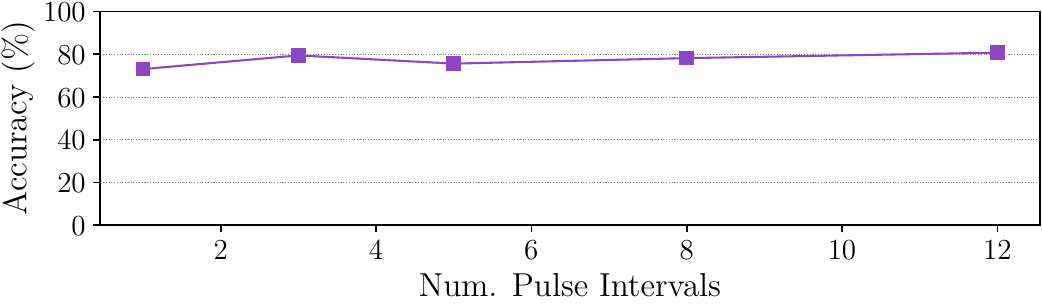}
    \vspace{1mm}
    \hrule
    \vspace{1mm}
    \caption{Accuracy on the Pima Indian Diabetes dataset using the square configuration with a spacing of 12$\mu m$. Changing the number of intervals used in the pulse parameterization impacts performance marginally. \sol{} uses three intervals for all pulses.}
    \label{fig:intervalablation}
\end{figure}

\begin{figure*}[t]
    \centering
    \subfloat{\includegraphics[width=0.32\textwidth]{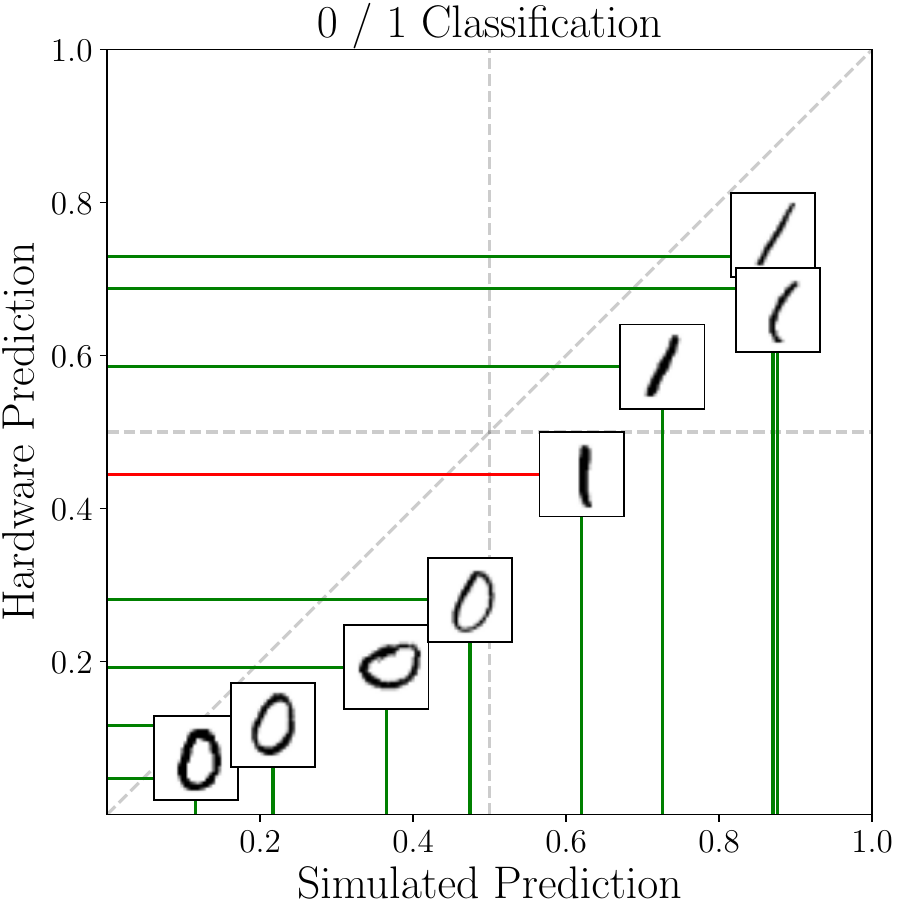}}
    \hfill
    \subfloat{\includegraphics[width=0.32\textwidth]{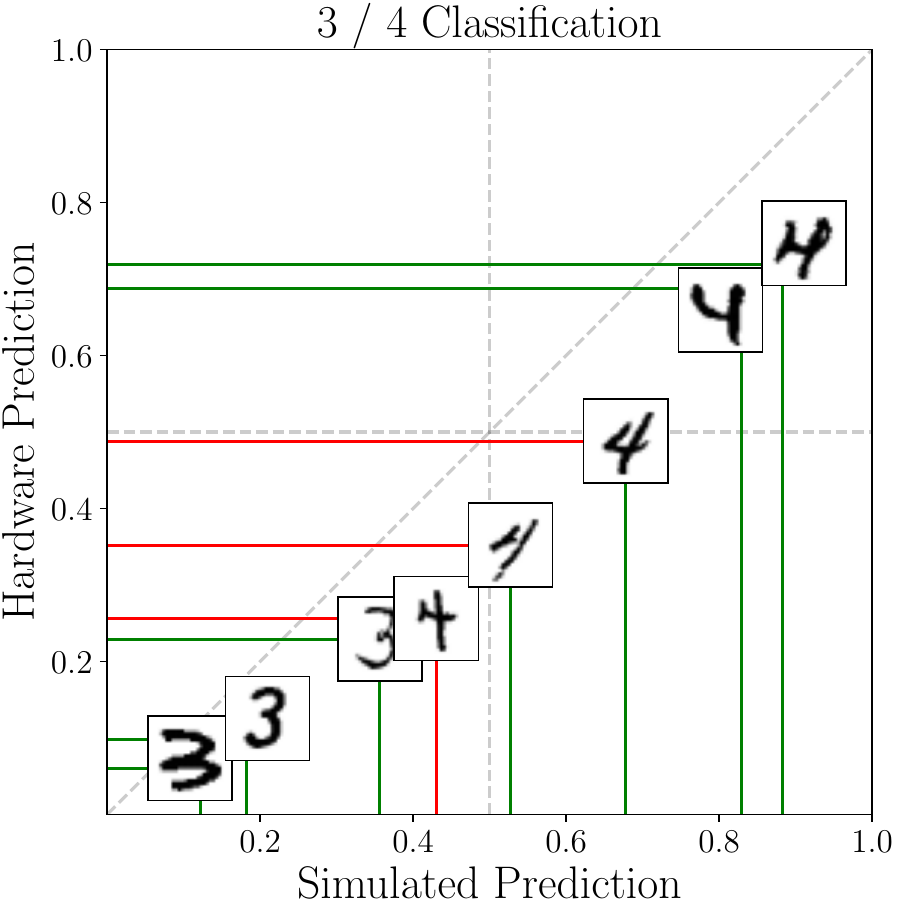}}
    \hfill
    \subfloat{\includegraphics[width=0.32\textwidth]{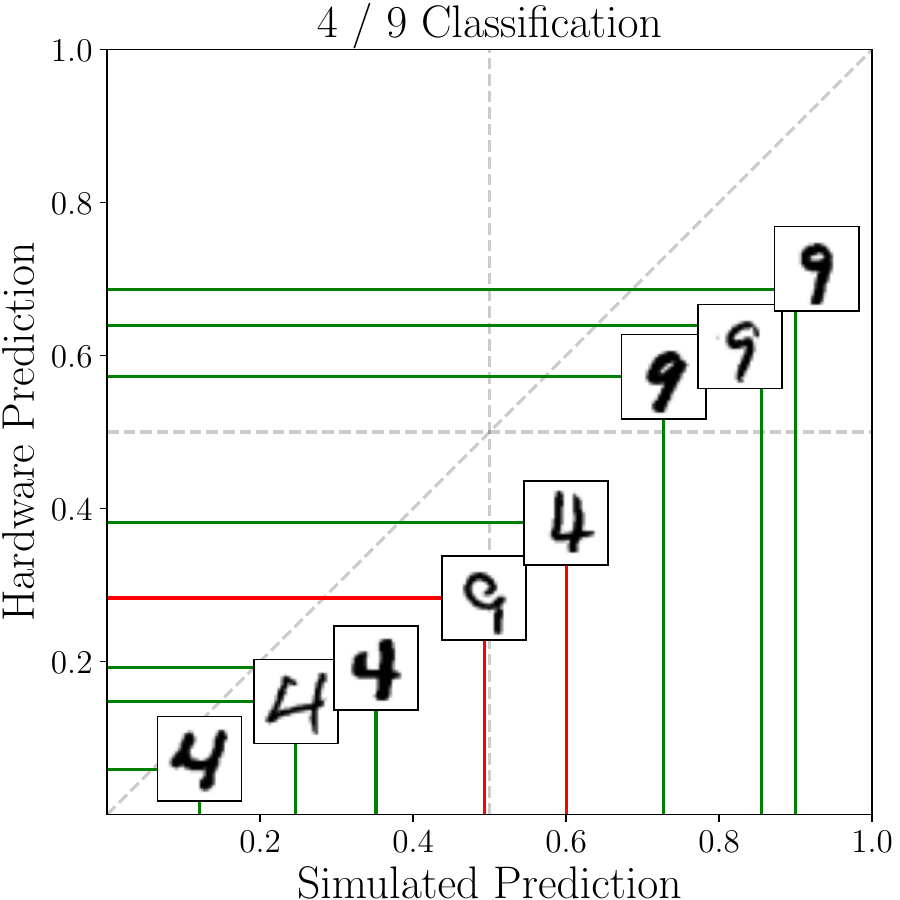}} \\
    \vspace{2mm}
    \subfloat{\includegraphics[width=0.32\textwidth]{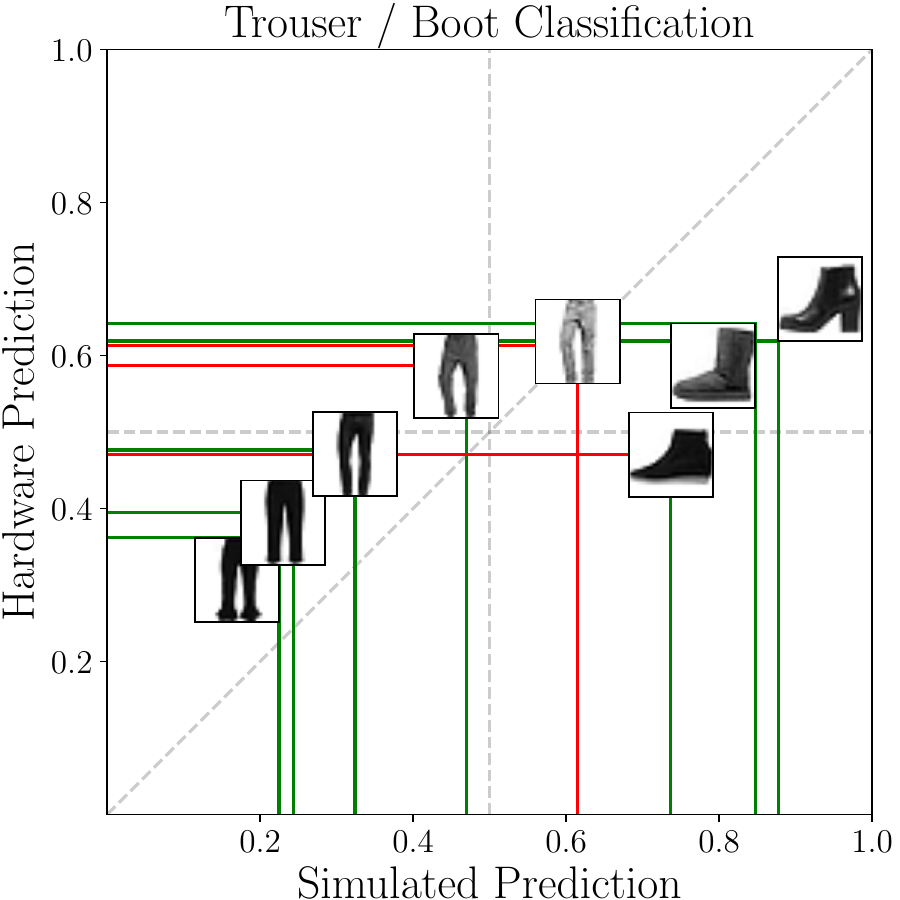}}
    \hfill
    \subfloat{\includegraphics[width=0.32\textwidth]{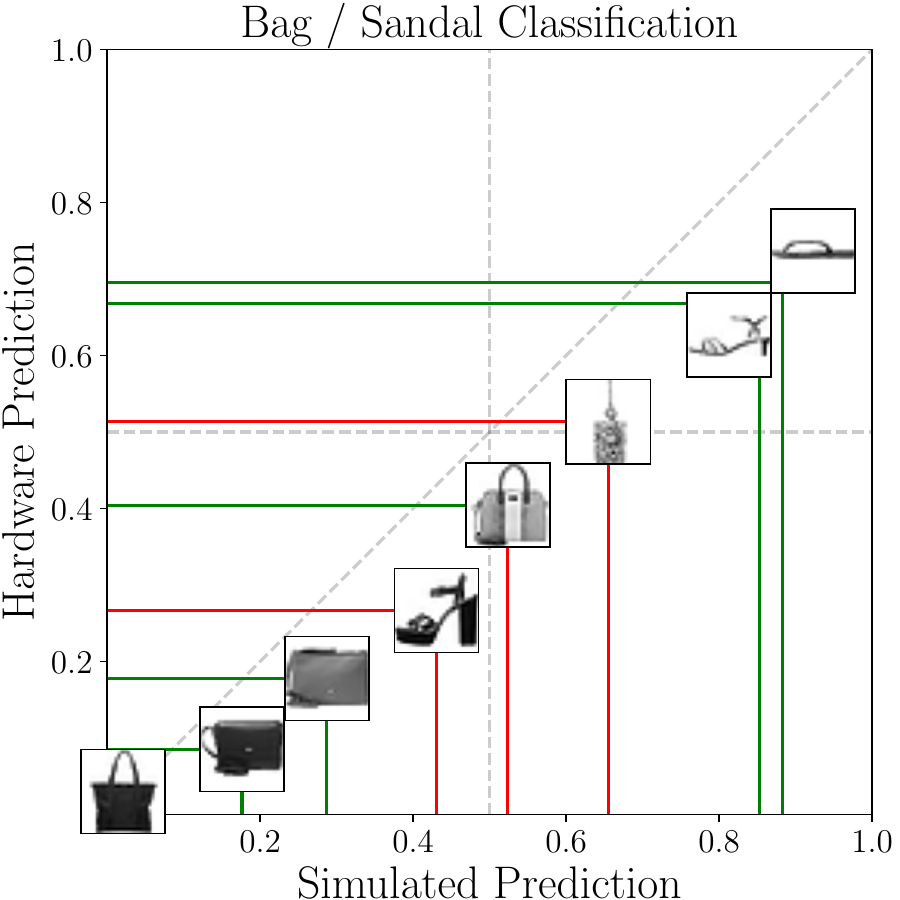}}
    \hfill
    \subfloat{\includegraphics[width=0.32\textwidth]{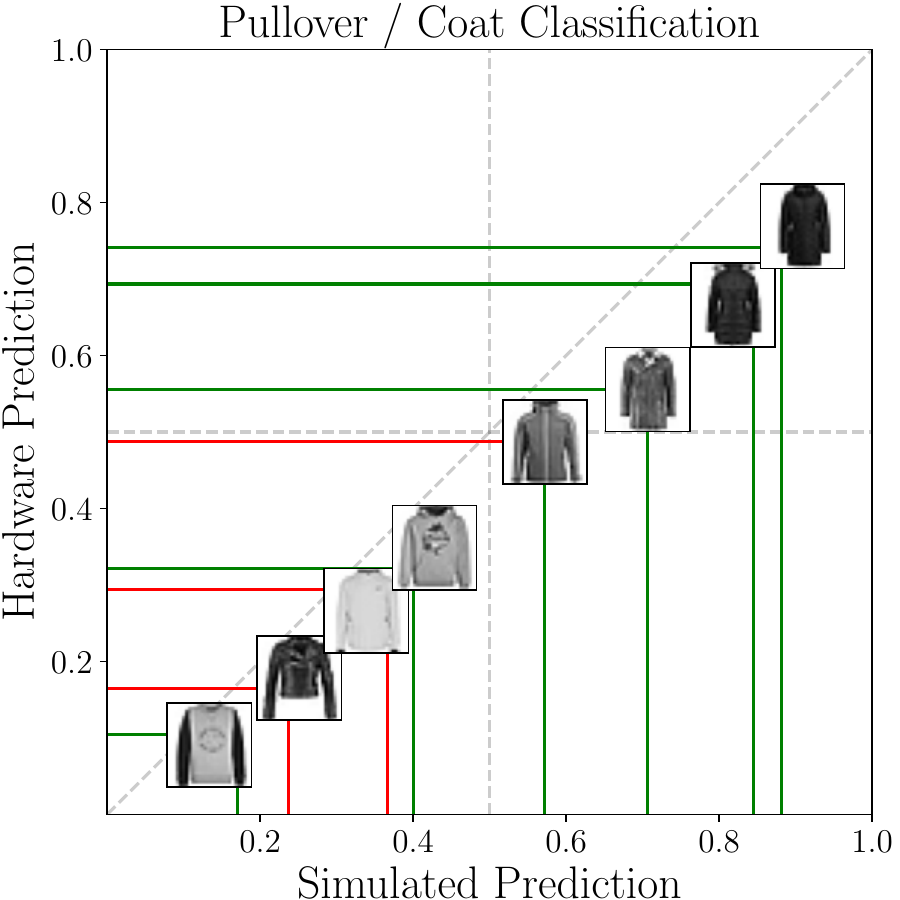}} \\
    \vspace{1.5mm}
    \hrule
    \vspace{-2.5mm}
    \caption{Prediction examples for MNIST and FashionMNIST datasets, using both ideal simulation and a real Rydberg atom computer. The vertical and horizontal lines indicate \sol{}'s prediction (measurement) for a given test sample, using ideal simulation and real quantum hardware, respectively. Green lines indicate a correct prediction, while red lines indicate an incorrect prediction. We see that hardware errors can flip decisions made on samples near a decision boundary (occurring at 0.5); however, many samples remain largely unaffected.}
\end{figure*}

\subsection{Impact of Pulse Intervals on \sol{}}

We investigate the importance of the number of pulse intervals in parameterizing the control pulses. We see a modest benefit to using three intervals when testing on the PID data, as shown in Figure~\ref{fig:intervalablation}. This relatively simple choice is also beneficial for simulation and training efficiency, as it was found that increasing the pulse interval count increased the simulation and training time. We observed similar trends for other datasets and classes.

\subsection{\sol{}'s Robustness on Noisy Hardware}

The previous evaluation relies on the ideal classical simulation of Rydberg atom computers. We now explore inference performance on noisy current-generation hardware.

\vspace{2mm}

\noindent\textbf{Real Hardware Executions.} We run inference tasks on the QuEra Aquila computer. Inference is performed on 8 samples from each of the MNIST and FashionMNIST classification tasks. Samples are chosen to have a variety of model prediction values (covering the spectrum of prediction values from 0 to 1) but are otherwise randomly selected. We find that \sol{}'s classifications are largely robust to hardware errors, but points close enough to a decision boundary (0.5) can have their predicted class flip due to hardware noise. The comparatively high errors of trouser/boot are likely due to it having the tightest geometry; the atoms being closer together allows for stronger interactions, which would allow for errors to accumulate faster. 


\vspace{2mm}

\noindent\textbf{Simulated Noise.} As it was prohibitively expensive to run a large number of samples on real hardware to perform a statistically significant analysis, we performed this analysis using simulated noise to emulate the noise characteristics of the Aquila computer. For neutral atom computers like Aquila, the primary source of noise is an error in the control pulses; this can be modeled by Gaussian perturbations to the laser pulses and atom positions, as is done in~\cite{lu2401digital} and~\cite{dibrita2024recon}. After doing so, \textit{we found no substantive difference in accuracy or F1 score when performing inference on a test set.} We did not plot these trends to avoid repetition. Generally, the accuracy and F1 score were within 1\% of the ideal simulation results, indicating that \sol{} has excellent resilience to noise as a large fraction of the samples do not cross the decision boundary.

\section{Conclusion}
\label{sec:conclusion}

In this work, we introduced \sol{}, a novel framework to leverage the natural dynamics of Rydberg atom quantum computers to implement ResNets. By aligning the continuous evolution of ResNets with the Hamiltonian control of Rydberg systems, \sol{} achieves an efficient, scalable approach to classification tasks, enabling linear scaling of input features with qubit count.

Our empirical results using simulations and real-machine runs on vision and healthcare datasets show that \sol{} not only overcomes the limitations of gate-based methods but also offers a powerful analog solution for implementing neural ODEs, performing well even in near-term noisy quantum environments. \textit{We envision that \sol{} will pave the way for further research into analog quantum systems, enabling advanced residual neural architectures, not just for computer vision applications but machine learning applications in general.} \noindent\textbf{\sol{}'s Code and Data:} {\small \url{https://github.com/positivetechnologylab/ResQ}}.

\section*{Acknowledgement}

This work was supported by Rice University, the Rice University George R. Brown School of Engineering and Computing, and the Rice University Department of Computer Science. This work was supported by the DOE Quantum Testbed Finder Award DE-SC0024301, the Ken Kennedy Institute, and Rice Quantum Initiative, which is part of the Smalley-Curl Institute. AWS Braket Cloud and the QuEra Aquila quantum computer were used for this work. The views expressed are those of the authors and do not reflect the official policy or position of AWS or QuEra teams.

\balance

{
    \small
    \bibliographystyle{ieeenat_fullname}
    \bibliography{main}
}

\end{document}